\documentclass[a4paper,12pt]{article}
\usepackage[top=2.5cm,bottom=2.5cm,left=2.5cm,right=2.5cm]{geometry}
\usepackage{amssymb}
\usepackage{graphicx}
\usepackage{amsmath,amsthm,amssymb,lineno}
\usepackage{latexsym}
\usepackage{CJK}
\usepackage{graphicx,booktabs,multirow}
\usepackage{appendix}

\newtheorem{theorem}{Theorem}
\newtheorem{lemma}[theorem]{Lemma}

\newtheorem{prop}[theorem]{Proposition}
\begin{document}

\title{Dimer-monomer Model on the Towers of Hanoi Graphs
\thanks{Project supported by Hunan Provincial Natural Science
Foundation of China (13JJ3053).}}
\author{Hanlin Chen, Renfang Wu, Guihua Huang, Hanyuan Deng\thanks{Corresponding author:
hydeng@hunnu.edu.cn}\\
{\footnotesize College of Mathematics and Computer Science,} \\
{\footnotesize Hunan Normal University, Changsha, Hunan 410081, P. R. China} \\
}

\date{}
\maketitle

\begin{abstract}

The number of dimer-monomers (matchings) of a graph $G$ is an
important graph parameter in statistical physics. Following recent
research, we study the asymptotic behavior of the number of
dimer-monomers $m(G)$ on the Towers of Hanoi graphs and another
variation of the Sierpi\'{n}ski graphs which is similar to the
Towers of Hanoi graphs, and derive the recursion relations for the
numbers of dimer-monomers. Upper and lower bounds for the entropy
per site, defined as $\mu_{G}=\lim_{v(G)\rightarrow\infty}\frac{\ln
m(G)}{v(G)}$, where $v(G)$ is the number of vertices in a graph $G$,
on these Sierpi\'{n}ski graphs are derived in terms of the numbers
at a certain stage. As the difference between these bounds converges
quickly to zero as the calculated stage increases, the numerical
value of the entropy can be evaluated with more than a hundred
significant figures accuracy.

\end{abstract}

\baselineskip=0.30in

\section{Introduction}\label{intro}

A classical model is the enumeration of the number of dimer-monomers
$m(G)$ on a graph $G$ in statistical physics (known as dimer-monomer
model; see \cite{ga,heli-1,heli-2,z}) and combinatorial chemistry
($m(G)$ is the Hosoya index; see \cite{gp}). In the dimer-monomer
model of statistical physics, each diatomic molecule is regarded as
a dimer which occupies two adjacent sites of the graph. The sites
that are not covered by any dimers are considered as occupied by
monomers. Since the general dimer-monomer problem was shown to be
computationally intractable \cite{je}, it is of interest to consider
the enumeration of the number of dimer-monomers for some classes of
graphs, and it has already been investigated extensively for trees,
hexagonal chains, grid graphs, and random graphs \cite{bm,je,ko}.

Recently, the enumeration of the number of dimer-monomers has been
considered in the physical literature for fractals and self-similar
graphs \cite{cc,tw-1,tw-2}. Following recent research on this graph
parameter in connection with self-similar, fractal-like graphs, we
will derive the recursion relations for the numbers of
dimer-monomers on the Towers of Hanoi graphs and another variation
of the Sierpi\'{n}ski graphs, and determine the entropies.

\section{Preliminaries}

We first recall some basic definitions about graphs. A graph
$G=(V,E)$ with vertex set $V$ and edge set $E$ is always supposed to
be undirected, without loops or multiple edges. Two vertices (sites)
$x$ and $y$ are adjacent if $xy$ is an edge (bond) in $E$. Let
$v(G)=|V|$ be the number of vertices and $e(G)=|E|$ the number of
edges in $G$. A matching (dimer-monomer) is a independent edge
subset. The number of matchings a graph $G$ is denoted by $m(G)$,
which is a parameter that is of relevance, among others, in
dimer-monomer model of statistical physics and combinatorial
chemistry. A parameter that is of particular interest is the
entropy
$$\mu_{v}=\lim_{v(G)\rightarrow\infty}\frac{\ln
m(G)}{v(G)}\,\,\,\,\,\,\,\,
\mbox{or}\,\,\,\,\,\,\,\,\mu_{e}=\lim_{e(G)\rightarrow\infty}\frac{\ln
m(G)}{e(G)}$$ On the analysis of large systems of recurrences, we
will see that the limit exists for the Towers of Hanoi graphs and
another variation of the Sierpi\'{n}ski graphs considered in this
paper.

There are many different approaches to construct self-similar
graphs. A construction that is specifically geared to be used in the
context of enumeration was described in \cite{tw-1}, it is no
restated and we will also make use of it here. Some examples can be
seen in \cite{tw-1,tw-3}.

{\bf Example 1}. The Tower of Hanoi graph, invented in 1883 by the
French mathematician Edouard Lucas, has become a classic example in
the analysis of algorithms and discrete mathematical structures
\cite{c-1,cn-1,h-4,h-5,hk-1,hp-1,ln-1,r-3}. It is a self-similar
graph and a slightly modified version of the Sierpi\'{n}ski graphs,
see Figure 1. The vertices of the graph $H_k$ in this sequence
correspond to all possible configurations of the game ¡°Towers of
Hanoi¡± with $k+1$ disks and three rods, whereas the edges describe
transitions between configurations. The Tower of Hanoi graphs can be
constructed following recursive-modular method. For $n=0$, $H_{0}$
is the complete graph $K_{3}$ (i.e., a 3-clique or triangle). For
$n\geq1$, $H_{n}$ is obtained from three graphs $H_{n-1}$ by
connecting them with three new edges.
  \begin{figure}[ht!]
\begin{center}
\includegraphics[width=12cm]{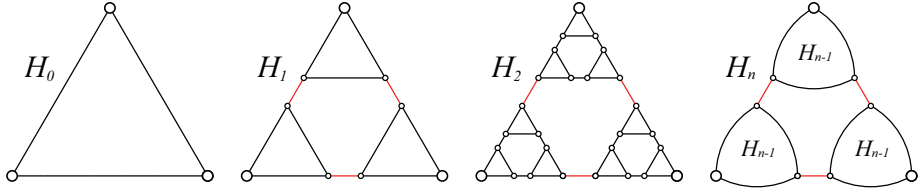}
\caption{The Tower of Hanoi graphs $H_{0},H_{1},H_{2}$ and the
construction of $H_{n}$.}
\end{center}
\end{figure}

{\bf Example 2}. The second example is another variation of the
Sierpi\'{n}ski graphs (similar to the Towers of Hanoi graphs) shown
in Figure 2, where $v(X_n)=\frac{7\times3^n-1}{2}$ and
$e(X_n)=\frac{5\times3^{n+1}-9}{2}$. Teufl and Wagner \cite{tw-3}
established a relation between the asymptotic growth of the number
of spanning trees and so-called (resistance) renormalization on
$X_n$.

\begin{figure}[ht!]
\begin{center}
\includegraphics[width=12cm]{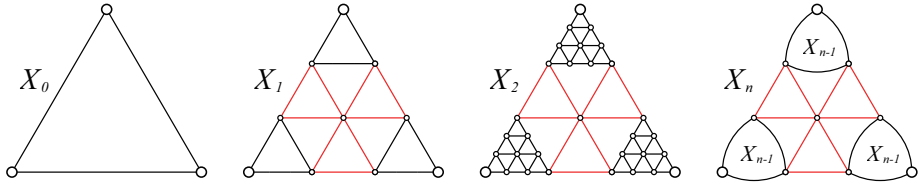}
\caption{The first three graphs $X_{0},X_{1},X_{2}$ and the construction of $X_{n}$.}
\end{center}
\end{figure}

\section{The number of Dimer-monomers on $H_{n}$}

In this section, we shall derive the recursion relations and the
entropy per site for the number of dimer-monomers on the
Tower of Hanoi graph $H_{n}$ in detail.

  \begin{figure}[ht!]
\begin{center}
\includegraphics[width=8cm]{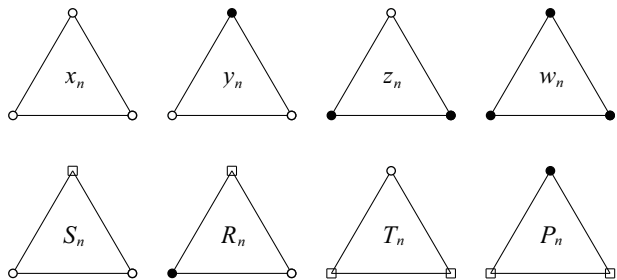}
\caption{Illustration for the configurations $x_n$, $y_n$, $z_n$,
$w_n$ and $S_{n}$, $R_{n}$, $T_{n}$, $P_{n}$, where each open
(solid) circle is occupied by a monomer (dimer), and each empty
squre is occupied by either a monomer or a dimer.}
\end{center}
\end{figure}

For the Tower of Hanoi graph $H_{n}$, $m_{n}$ is its number of
dimer-monomers, $x_{n}$ is its number of dimer-monomers such that
all three outmost vertices are occupied by monomers, $y_{n}$ is its
number of dimer-monomers such that only one specified vertex of
three outmost vertices is occupied by a dimer and the other two
outmost vertices are occupied by monomers, $z_{n}$ is its number of
dimer-monomers such that exact two specified vertices of the three
outmost vertices are occupied by dimers and the other one is
occupied by a monomer, $w_{n}$ is its number of dimer-monomers such
that all three outmost vertices are occupied by dimers. They are
illustrated in Figure 3, where only the outmost vertices are shown
and a solid circle is occupied by a dimer and an open circle is
occupied by a monomer. Because of rotational symmetry, there are
three possible $y_{n}$ and three possible $z_{n}$ such that
$$m_{n}=x_{n}+3y_{n}+3z_{n}+w_{n}$$
and $x_{0}=1$, $y_{0}=0$, $z_{0}=1$, $w_{0}=0$,
$m_{0}=x_{0}+3y_{0}+3z_{0}+w_{0}=4$.

Thus, we have reduced the problem of determining $m_{n}$ to finding
the four quantities $x_n$, $y_n$, $z_n$ and $w_n$, which is the
theme of what follows. To find these intermediate variables, we
define four additional quantities $S_n, R_n, T_n$ and $P_n$, see
Figure 3. Let $S_n$ be the number of dimer-monomer arrangements on
$H_{n}$ such that one outmost vertex is occupied by either a monomer
or a dimer while the other two outmost vertices are occupied by
monomers.
Similarly, one can define the quantities $R_n$, $T_{n}$
and $P_{n}$. By definitions, we have $S_{n}=x_{n}+y_{n}$,
$R_{n}=y_{n}+z_{n}$, $T_{n}=x_{n}+2y_{n}+z_{n}$ and
$P_{n}=y_{n}+2z_{n}+w_{n}$.

\begin{lemma} \label{l1}
For any positive integer $n$,
\begin{align}
x_{n+1}=&8x^{3}_{n}+24x^{2}_{n}y_{n}+6x^{2}_{n}z_{n}+30x_{n}y^{2}_{n}
+18x_{n}y_{n}z_{n}\nonumber
\\
&+3x_{n}z^{2}_{n}+14y^{3}_{n}+15y^{2}_{n}z_{n}+6y_{n}z^{2}_{n}+z^{3}_{n},
\end{align}
\begin{align}
y_{n+1}=&8x^{2}_{n}y_{n}+8x^{2}_{n}z_{n}+2x^{2}_{n}w_{n}+16x_{n}y^{2}_{n}+24x_{n}y_{n}z_{n}
+6x_{n}y_{n}w_{n}+6x_{n}z^{2}_{n}+2x_{n}z_{n}w_{n}\nonumber
\\
&+10y^{3}_{n} +20y^{2}_{n}z_{n}+5y^{2}_{n}w_{n}
+11y_{n}z^{2}_{n}+4y_{n}z_{n}w_{n}+2z^{3}_{n}+z^{2}_{n}w_{n},
\end{align}
\begin{align}
z_{n+1}=&8x_{n}y^{2}_{n}+16x_{n}y_{n}z_{n}+4x_{n}y_{n}w_{n}+10x_{n}z^{2}_{n}
+6x_{n}z_{n}w_{n}+x_{n}w^{2}_{n}+8y^{3}_{n}+22y^{2}_{n}z_{n}\nonumber
\\
&+6y^{2}_{n}w_{n}
+20y_{n}z^{2}_{n}+12y_{n}z_{n}w_{n}+2y_{n}w^{2}_{n}+5z^{3}_{n}+4z^{2}_{n}w_{n}+z_{n}w^{2}_{n},
\end{align}
\begin{align}
w_{n+1}=&8y^{3}_{n}+24y^{2}_{n}z_{n}+6y^{2}_{n}w_{n}+30y_{n}z^2_{n}+18y_{n}z_{n}w_{n}\nonumber
\\
&+3y_{n}w^{2}_{n}+ 14z^{3}_{n}+15z^{2}w_{n}+6z_{n}w^{2}_{n}+w^{3}_{n}
\end{align}
with the initial values $x_{1}=18$, $y_{1}=16$, $z_{1}=15$ and
$w_{1}=14$ at stage $n=1$.
\end{lemma}
{\bf Proof}. Note that $H_{n+1}$ is obtained from three copies of
graph $H_{n}$ connected by three new edges, see Figure 1. To
simplify our calculation, we introduce the following the classes of
dimer-monomers in $H_{n+1}$ in terms of the subset of three new
edges, which can be contained in the matching (i.e. two terminal
vertices of the new edges in the subset are occupied by dimers). The
number $x_{n+1}$ consists of (i) one configuration where none of new
edges are contained in the matching; (ii) three configurations where
one of new edges is contained in the matching; (iii) three
configurations where two of new edges are contained in the matching;
(iv) one configuration where all of the three new
edges are contained in the matching,
as illustrated in Figure 4.
\begin{figure}[ht!]
\begin{center}
\includegraphics[width=10cm]{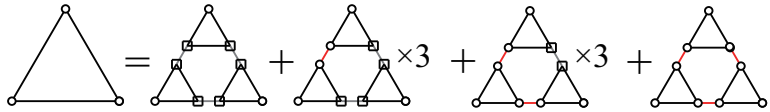}
\caption{Illustration for the expression of $x_{n+1}$.The
multiplication of three on the right-hand-side corresponds to the
three possible orientations of $H_{n+1}$. Here, we just present the
status of three copies of graph $H_n$, which contributed to
$H_{n+1}$, and the two terminal vertices of new adding edge are
omit, when it occupied by two dimers(i.e.the edge is covered by a matching), we use a red line instead,
otherwise, a gray line.}
\end{center}
\end{figure}

\begin{align}
x_{n+1}&=T_{n}^{3}+3S_{n}^{2}T_{n}+3x_{n}S_{n}^{2}+x_{n}^{3}\nonumber\\
&=(x_{n}+2y_n+z_n)^{3}+3(x_n+z_n)^{2}(x_{n}+2y_n+z_n)+3x_{n}(x_n+y_n)^{2}+x_{n}^{3}\nonumber\\
&=8x^{3}_{n}+24x^{2}_{n}y_{n}+6x^{2}_{n}z_{n}+30x_{n}y^{2}_{n}
+18x_{n}y_{n}z_{n}+3x_{n}z^{2}_{n}+14y^{3}_{n}+15y^{2}_{n}z_{n}+6y_{n}z^{2}_{n}+z^{3}_{n}.\nonumber
\end{align}

Similarly, the expressions of $y_{n+1}, z_{n+1}$ and $w_{n+1}$ can
be obtained with appropriate configurations of its three $H_{n}$ as
illustrated in Figures 5-7. \hfill $\Box$ \\

  \begin{figure}[ht!]
\begin{center}
\includegraphics[width=14cm]{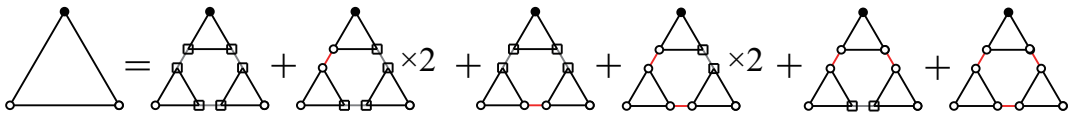}
\caption{Illustration for the expression of $y_{n+1}$. The
multiplication of two on the right-hand-side corresponds to the
reflection symmetry with respect to the central vertical axis.}
\end{center}
\end{figure}

  \begin{figure}[ht!]
\begin{center}
\includegraphics[width=14cm]{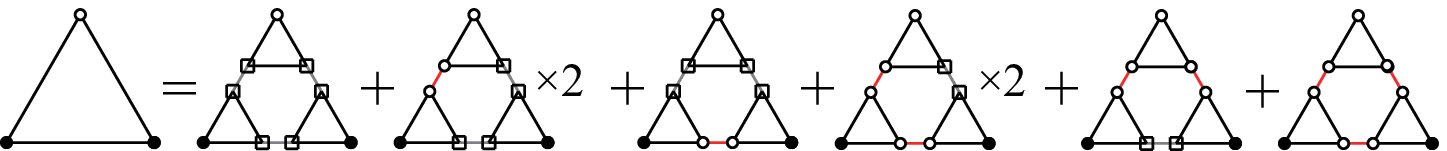}
\caption{Illustration for the expression of $z_{n+1}$. The
multiplication of two on the right-hand-side corresponds to the
reflection symmetry with respect to the central vertical axis.}
\end{center}
\end{figure}

  \begin{figure}[ht!]
\begin{center}
\includegraphics[width=10cm]{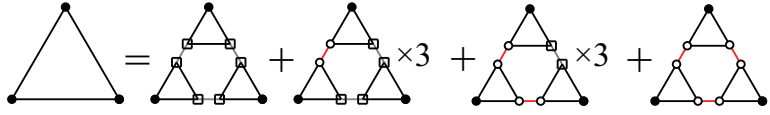}
\caption{Illustration for the expression of $w_{n+1}$. The
multiplication of three on the right-hand-side corresponds to the
three possible orientations of $H_{n+1}$.}
\end{center}
\end{figure}

In the following, we will estimate the value
$\mu_{G}=\lim_{v\rightarrow\infty}\frac{\ln m(G)}{v(G)}$ of the
entropy for the Tower of Hanoi graph $G=H_{n}$.
The values of $x_{n}, y_{n}, z_{n}, w_{n}$ for small $n$ are listed
in Table 1 by Eqs. (1)-(4), and grow exponentially. For the Tower of
Hanoi graph $H_{n}$, we define the ratios
\begin{align}\label{eq-5}
\alpha_n=\frac{y_{n}}{x_{n}},\,\,\,\,\,
\beta_n=\frac{z_{n}}{y_{n}},\,\,\,\,\,
\gamma_n=\frac{w_{n}}{z_{n}}
\end{align}
where $n$ is a positive integer. Their values for small $n$ are
listed in Table 2. From the initial values of $x_{n}, y_{n}, z_{n},
w_{n}$, it is easy to see that $x_{n}>y_{n}>z_{n}>w_{n}$ and it holds for all positive
integer $n$ proved easily by induction. It follows that $\alpha_{n},
\beta_{n}, \gamma_{n}\in (0,1)$.

\begin{table*}[!t]
\centering
\caption{The first few values of $x_{n}, y_{n}, z_{n}, w_{n}$.}
\label{T1}
\begin{tabular}{ccccc}
\toprule[1pt]
$n$&0&1&2&3\\
\midrule
$x_{n}$&1&18&568,301&18,782,596,680,434,061,312\\

$y_{n}$&0&16&521,504&17,236,435,531,779,805,184\\

$z_{n}$&1&15&478,579&15,817,552,541,478,490,112\\

$w_{n}$&0&14&439,204&14,515,470,321,889,908,736\\

\bottomrule[1pt]
\end{tabular}
\end{table*}

\begin{table*}[!t]
\centering
\caption{The first few values of $\alpha_n, \beta_n, \gamma_n$.}
\label{T1}
\begin{tabular}{cccc}
\toprule[1pt]
$n$&1&2&3\\
\midrule
$\alpha_n$& 0.888888888888889&0.917654552781009&0.9176811824818183\\

$\beta_n$& 0.9375&0.917689988955022&0.9176811825342172\\

$\gamma_n$&0.933333333333333& 0.917725182258311&0.9176811825866157\\
\bottomrule[1pt]
\end{tabular}
\end{table*}

\begin{lemma}\label{l2}
For any positive integer $n(n\geq2)$, the ratios satisfy
$$\alpha_{n}<\beta_{n}<\gamma_{n}.$$
When $n$ increases, the ratio $\alpha_{n}$ increases monotonically
while $\gamma_{n}$ decreases monotonically. The three ratios in the
large $n$ limit are equal to each other
$$\lim_{n\rightarrow\infty}\alpha_{n}=\lim_{n\rightarrow\infty}\beta_{n}=\lim_{n\rightarrow\infty}\gamma_{n}.$$
\end{lemma}
{\bf Proof}. By the definition of $\alpha_n, \beta_n, \gamma_n$, we
have
$$\alpha_{n+1}=\alpha_{n}\frac{B_{n}}{A_{n}},~\beta_{n+1}=\alpha_{n}\frac{C_{n}}{B_{n}},~\gamma_{n+1}=\alpha_{n}\frac{D_{n}}{C_{n}}$$
for a positive integer $n$, where

$A_{n}=8+24\alpha_{n}+6\alpha_{n}\beta_{n}+30\alpha_{n}^{2}+18\alpha_{n}^{2}\beta_{n}+3\alpha_{n}^{2}\beta_{n}^{2}
+14\alpha_{n}^{3}+15\alpha_{n}^{3}\beta_{n}+6\alpha_{n}^{3}\beta_{n}^{2}+\alpha_{n}^{3}\beta_{n}^{3},$

$B_{n}=8+8\beta_{n}+2\beta_{n}\gamma_{n}+16\alpha_{n}+24\alpha_{n}\beta_{n}+6\alpha_{n}\beta_{n}\gamma_{n}+6\alpha_{n}\beta_{n}^{2}
+2\alpha_{n}\beta_{n}^{2}\gamma_{n}+10\alpha_{n}^{2}+20\alpha_{n}^{2}\beta_{n}+5\alpha_{n}^{2}\beta_{n}\gamma_{n}
+11\alpha_{n}^{2}\beta_{n}^{2}+4\alpha_{n}^{2}\beta_{n}^{2}\gamma_{n}+2\alpha_{n}^{2}\beta_{n}^{3}+\alpha_{n}^{2}\beta_{n}^{3}\gamma_{n},$

$C_{n}=8+16\beta_{n}+4\beta_{n}\gamma_{n}+10\beta_{n}^{2}+6\beta_{n}^{2}\gamma_{n}+\beta_{n}^{2}\gamma_{n}^{2}+8\alpha_{n}
+22\alpha_{n}\beta_{n}+6\alpha_{n}\beta_{n}\gamma_{n}+20\alpha_{n}\beta_{n}^{2}+12\alpha_{n}\beta_{n}^{2}\gamma_{n}
+2\alpha_{n}\beta_{n}^{2}\gamma_{n}^{2}+5\alpha_{n}\beta_{n}^{3}+4\alpha_{n}\beta_{n}^{3}\gamma_{n}+\alpha_{n}\beta_{n}^{3}\gamma_{n}^{2},$

$D_{n}=8+24\beta_{n}+6\beta_{n}\gamma_{n}+30\beta_{n}^{2}+18\beta_{n}^{2}\gamma_{n}+3\beta_{n}^{2}\gamma_{n}^{2}+14\beta_{n}^{3}+
15\beta_{n}^{3}\gamma_{n}+6\beta_{n}^{3}\gamma_{n}^{2}+\beta_{n}^{3}\gamma_{n}^{3}.$

In the following, we show that
$\frac{1}{2}<\alpha_{n}<\beta_n<\gamma_{n}<1$ by induction on $n$.
It is true for $n=2, 3$ from Table 2. Suppose that
$\frac{1}{2}<\alpha_{n}<\beta_n<\gamma_{n}<1$ for $n\geq 3$.

Let $\varepsilon_{n}=\alpha_{n}-\gamma_{n}$. Then
$\varepsilon_{n}>\alpha_{n}-\beta_{n}$,
$\varepsilon_{n}>\beta_{n}-\gamma_{n}$ and
$\varepsilon_{n}\in(0,\frac{1}{2})$. Now,
\begin{align}
\alpha_{n+1}-\alpha_{n}&=\alpha_{n}\frac{B_{n}}{A_{n}}-\alpha_{n}=\frac{\alpha_n(B_n-A_n)}{A_n}
\nonumber \\
&=\frac{\alpha_{n}}{A_{n}}[(8+20\alpha_{n}+6\alpha_{n}\beta_{n}+14\alpha_{n}^{2}+10\alpha_{n}^{2}\beta_{n}+
2\alpha_{n}^{2}\beta_{n}^{2})(\beta_{n}-\alpha_{n})
+(2\beta_{n}+6\alpha_{n}\beta_{n}\nonumber\\&+2\alpha_{n}\beta_{n}^{2}+5\alpha_{n}^{2}\beta_{n}+4\alpha_{n}^{2}\beta_{n}^{2}
+\alpha_{n}^{2}\beta_{n}^{3})(\gamma_{n}-\alpha_{n})]\nonumber\\
&>0\nonumber\\
\beta_{n+1}-\alpha_{n+1}&=\frac{\alpha_{n}(A_{n}C_{n}-B_{n}^{2})}{A_{n}B_{n}}>0\nonumber
\end{align}
where

$
A_{n}C_{n}-B_{n}^{2}=(16+32\alpha_{n}+12\alpha_{n}^{2}+8\alpha_{n}\beta_{n}+6\alpha_{n}^{2}\beta_{n}^{2}+
2\alpha_{n}\beta_{n}^{3}+\alpha_{n}^2\beta_{n}^3\gamma_{n})(\beta_{n}-\alpha_{n})^{2}
+(16\beta_{n}+44\alpha_{n}\beta_{n}+4\beta_{n}^{2}+28\alpha_{n}^{2}\beta_{n}+12\alpha_{n}\beta_{n}^{2}
+16\alpha_{n}^{2}\beta_{n}^{2}+4\alpha_{n}\beta_{n}^{3}+8\alpha_{n}^{2}\beta_{n}^{3})(\beta_{n}-
\alpha_{n})(\gamma_{n}-\alpha_{n})+(4\beta_{n}^{2}+22\alpha_{n}^{2}\beta_{n}^{2}+6\alpha_{n}\beta_{n}^{3}+
14\alpha_{n}^{3}\beta_{n}^{2}+14\alpha_{n}^{2}\beta_{n}^{3}+13\alpha_{n}^{3}\beta_{n}^{3}+\alpha_{n}^{2}\beta_{n}^{4}
+2\alpha_{n}^{3}\beta_{n}^{4})(\gamma_{n}-\alpha_{n})(\gamma_{n}-\beta_{n})+(4\alpha_{n}\beta_{n}+
16\alpha_{n}^{2}\beta_{n}+16\alpha_{n}^{3}\beta_{n}+14\alpha_{n}^{2}\beta_{n}^{2}+2\alpha_{n}\beta_{n}^{3}+
8\alpha_{n}^{3}\beta_{n}^{2})(\beta_{n}-\alpha_{n})(\gamma_{n}-\beta_{n})+16\alpha_{n}\beta_{n}^{2}(\gamma_{n}-\alpha_{n})^{2}
+(3\alpha_{n}^{4}\beta_{n}^{2}+4\alpha_{n}^{4}\beta_{n}^{3}+\alpha_{n}^{4}\beta_{n}^{4})(\gamma_{n}-\beta_{n})^{2}
$

$
A_{n}B_{n}>\alpha_{n}\{16+32\alpha_{n}+16\beta_{n}+56\alpha_{n}\beta_{n}
+28\alpha_{n}\beta_{n}^{2}+44\alpha_{n}^{2}\beta_{n}+14\alpha_{n}\beta_{n}^{3}+16\alpha_{n}^{3}\beta_{n}+12\alpha_{n}^{2}
+8\beta_{n}^{2}+58\alpha_{n}^{2}\beta_{n}^{2}+22\alpha_{n}^{2}\beta_{n}^{3}+22\alpha_{n}^{3}\beta_{n}^{2}+\alpha_{n}^{2}\beta_{n}^{4}
+13\alpha_{n}^{3}\beta_{n}^{3}+3\alpha_{n}^{4}\beta_{n}^{2}+2\alpha_{n}^{3}\beta_{n}^{4}+4\alpha_{n}^{4}\beta_{n}^{3}+
\alpha_{n}^{4}\beta_{n}^{4}+\alpha_{n}^{2}\beta_{n}^{3}\gamma_{n}\}
$

And

$
\beta_{n+1}-\alpha_{n+1}=\frac{\alpha_{n}(A_{n}C_{n}-B_{n}^{2})}{A_{n}B_{n}}
<\frac{\varepsilon^{2}_{n}}{A_{n}B_{n}}\{\alpha_{n}[16+32\alpha_{n}+16\beta_{n}+56\alpha_{n}\beta_{n}
+28\alpha_{n}\beta_{n}^{2}+44\alpha_{n}^{2}\beta_{n}+14\alpha_{n}\beta_{n}^{3}+16\alpha_{n}^{3}\beta_{n}+12\alpha_{n}^{2}
+8\beta_{n}^{2}+58\alpha_{n}^{2}\beta_{n}^{2}+22\alpha_{n}^{2}\beta_{n}^{3}+22\alpha_{n}^{3}\beta_{n}^{2}+\alpha_{n}^{2}\beta_{n}^{4}
+13\alpha_{n}^{3}\beta_{n}^{3}+3\alpha_{n}^{4}\beta_{n}^{2}+2\alpha_{n}^{3}\beta_{n}^{4}+4\alpha_{n}^{4}\beta_{n}^{3}+
\alpha_{n}^{4}\beta_{n}^{4}+\alpha_{n}^{2}\beta_{n}^{3}\gamma_{n}]\}<\varepsilon^{2}_{n}
$\\
since $\varepsilon_{n}>\alpha_{n}-\beta_{n}$ and
$\varepsilon_{n}>\beta_{n}-\gamma_{n}$.

Similarly, we have
$$\gamma_{n+1}-\beta_{n+1}=\frac{\alpha_{n}(B_{n}D_{n}-C_{n}^{2})}{B_{n}C_{n}}>0\,\,\,\,\,\,\mbox{and}
\,\,\,\,\,\, \gamma_{n+1}-\beta_{n+1}<\varepsilon^{2}_{n}.$$
\begin{align}
\gamma_{n}-\gamma_{n+1}=&\frac{1}{C_n}(\gamma_nC_n-\alpha_nD_n)\nonumber\\
=&\frac{1}{C_{n}}[(8+16\beta_{n}+10\beta_{n}^{2}+4\beta_{n}\gamma_{n}+6\beta_{n}^{2}\gamma_{n}
+\beta_{n}^{2}\gamma_{n}^{2})(\gamma_{n}-\alpha_{n})+\nonumber\\
&
(8\alpha_{n}+20\alpha_{n}\beta_{n}+14\alpha_{n}\beta_{n}^{2}+6\alpha_{n}\beta_{n}\gamma_{n}+10\alpha_{n}\beta_{n}^{2}\gamma_{n}+2\alpha_{n}\beta_{n}^{2}\gamma_{n}^{2})(\gamma_{n}-\beta_{n})]>0\nonumber
\end{align}

So, we have (i) $\alpha_{n+1}-\alpha_{n}>0$, (ii)
$0<\beta_{n+1}-\alpha_{n+1}<\varepsilon^{2}_{n}$, (iii)
$0<\gamma_{n+1}-\beta_{n+1}<\varepsilon^{2}_{n}$ and (iv)
$\gamma_{n}-\gamma_{n+1}>0$.

From (ii) and (iii), we can obtain that
$\varepsilon_{n+1}=\gamma_{n+1}-\alpha_{n+1}<2\varepsilon^{2}_{n}<\frac{1}{2}$
for all positive integer $n$ by induction. It follows that for any
positive integer $k\leq n$,
$$\varepsilon_{n}<2\varepsilon^{2}_{n-1}<2[2\varepsilon^{2}_{n-2}]^{2}<...<\frac{1}{2}[2\varepsilon_{k}]^{2^{n-k}}$$
Since $\varepsilon_{k}\in (0,\frac{1}{2})$ for any positive integer
$k$, we have that the values of $\alpha_{n}, \beta_{n}, \gamma_{n}$
are close to each other when $n$ becomes large. \hfill$\Box$

Numerically, we can find
$$\lim_{n\rightarrow\infty}\alpha_{n}=\lim_{n\rightarrow\infty}\beta_{n}
=\lim_{n\rightarrow\infty}\gamma_{n}=0.9176811825212464\cdots$$ From
the lemmas above, we get the bounds for the number of
dimer-monomers.

\begin{lemma}\label{13}
For any positive integer $k\leq n$,
$$x_{k}^{3^{n-k}}(\alpha_{k}^{2}+2\alpha_{k}+2)^{\frac{3(3^{n-k}-1)}{2}}(1+\alpha_{n})^{3}<m(H_{n})
<x_{k}^{3^{n-k}}(\beta_{k}^{2}+2\beta_{k}+2)^{\frac{3(3^{n-k}-1)}{2}}(1+\gamma_{n})^{3}.$$
\end{lemma}
{\bf Proof}. By Lemmas 1 and 2 and the definition of $\alpha_n,
\beta_n, \gamma_n$, we have
\begin{align}
x_{n}=&x^{3}_{n-1}(8+24\alpha_{n}+6\alpha_{n}\beta_{n}+30\alpha_{n}^{2}+18\alpha_{n}^{2}\beta_{n}+3\alpha_{n}^{2}\beta_{n}^{2}
+14\alpha_{n}^{3}+15\alpha_{n}^{3}\beta_{n}+6\alpha_{n}^{3}\beta_{n}^{2}+\alpha_{n}^{3}\beta_{n}^{3})\nonumber\\
<&x^{3}_{n-1}(8+24\beta_{n-1}+36\beta_{n-1}^{2}+32\beta_{n-1}^{3}+18\beta_{n-1}^{4}+6\beta_{n-1}^{5}+\beta_{n-1}^{6})\nonumber\\
=&[x_{n-1}(\beta_{n-1}^{2}+2\beta_{n-1}+2)]^{3}<[x^{3}_{n-2}(\beta_{n-2}^{2}+2\beta_{n-2}+2)^{3}]^{3}(\beta_{n-1}^{2}+2\beta_{n-1}+2)^{3}\nonumber\\
<&x^{3^2}_{n-2}(\beta_{n-2}^{2}+2\beta_{n-2}+2)^{3^2+3^1}\nonumber\\
<&\cdots<x_{k}^{3^{n-k}}(\beta_{k}^{2}+2\beta_{k}+2)^{\frac{3(3^{n-k}-1)}{2}}\nonumber
\end{align}
And
\begin{align}
m(H_{n})=&x_{n}+3y_{n}+3z_{n}+w_{n}
=x_{n}(1+3\alpha_{n}+3\alpha_{n}\beta_{n}+\alpha_{n}\beta_{n}\gamma_{n})\nonumber\\
<&x_{n}(1+3\gamma_{n}+3\gamma^{2}_{n}+\gamma^{3}_{n})
=x_{n}(1+\gamma_{n})^{3}
<x_{k}^{3^{n-k}}(\beta_{k}^{2}+2\beta_{k}+2)^{\frac{3(3^{n-k}-1)}{2}}(1+\gamma_{n})^{3}\nonumber
\end{align}

Similarly, the lower bound for $m(H_{n})$ can be derived.
\hfill$\Box$

\begin{lemma}\label{14}
The entropy for the number of dimer-monomers in
$H_{n}$ is bonded by
$$\frac{\ln x_{k}}{3^{k+1}}+\frac{\ln(\alpha_{k}^{2}+2\alpha_{k}+2)}{2\times3^{k}}\leq \mu_{H}
\leq \frac{\ln x_{k}}{3^{k+1}}+\frac{\ln
(\beta_{k}^{2}+2\beta_{k}+2)}{2\times3^{k}}$$ where $k$ is a
positive integer and
$\mu_{H}=\lim_{v(H_n)\rightarrow\infty}\frac{\ln m(H_{n})}{v(H_n)}$.
\end{lemma}
{\bf Proof}. Note that the number of vertices of $H_{n}$ is
$v(H_n)=3^{n+1}$, by Lemma \ref{13}, we have
$$\frac{\ln m(H_n)}{v(H_n)}<\frac{\ln x_{k}}{3^{k+1}}+\frac{\ln
(\beta_{k}^{2}+2\beta_{k}+2)}{2\times
3^{k}}-\frac{\ln(\beta_{k}^{2}+2\beta_{k}+2)}{2\times
3^{n}}+\frac{\ln (1+\alpha_{n})}{3^{n}}$$ and
$$\frac{\ln m(H_{n})}{v(H_n)}>\frac{\ln x_{k}}{3^{k+1}}+\frac{\ln
(\alpha_{k}^{2}+2\alpha_{k}+2)}{2\times
3^{k}}-\frac{\ln(\alpha_{k}^{2}+2\alpha_{k}+2)}{2\times
3^{n}}+\frac{\ln (1+\beta_{n})}{3^{n}}$$

So, the bounds for $\mu_{H}=\lim_{v(H_n)\rightarrow\infty}\frac{\ln
m(H_{n})}{v(H_n)}$ follow. \hfill$\Box$\\

As $k$ increases, the difference between the upper and lower bounder
in Lemma \ref{14} becomes small and the convergence is rapid.
The numerical value of $\mu_{H}$ can be obtained with more than a
hundred significant figures accurate when $k$ is equal to seven.
It is too lengthy to be included here and is available from the
authors on request.
\begin{prop}
The entropy per site for the number of
dimer-monomers of the Hanoi Tower graph $H_{n}$ in the large $n$
limit is $\mu_{H}=0.5764643016505283756\cdots$.
\end{prop}

\section{The number of Dimer-monomers on $X_{n}$}

In this section, we find that the method given in the previous
section can be also applied to the number of dimer-monomers on
another variation of the Sierpi\'{n}ski graphs $X_{n}$, see Figure
2. For $X_{n}$, the numbers of vertices and edges are given by
$$v(X_{n})=\frac{7\times3^{n}-1}{2},~~~e(X_{n})=\frac{5\times3^{n+1}-9}{2}.$$
There are three outmost vertices of $X_{n}$ with degree two,
$\frac{3^{n}-1}{2}$ vertices with degree six and $3(3^{n}-1)$
vertices with degree four.
The values of $x_{n}, y_{n}, z_{n}, w_{n}$ on $X_n$ for small $n$
are listed in Table 3, and also grow exponentially, without simple integer
factorizations. The recursion relations  are lengthy and given in the appendix.
Similarly, for the
sequences of the ratio defined in Eq.(\ref{eq-5}) on $X_n$,
$\{\alpha_{n}\}_{n=1}^{\infty}$ increases monotonically and
$\{\gamma_{n}\}_{n=1}^{\infty}$ decreases monotonically with
$1/2<\alpha_{n}<\beta_{n}<\gamma_{n}<1,$ and the limits
$\alpha_{n},\beta_{n}$, and $\gamma_{n}$ are equal to each other,
the same as the results for $H_{n}$ in lemma 2. And the first few values of
$\alpha_{n},\beta_{n},\gamma_{n}$ for $X_n$ are listed in Table 4.

\begin{table*}[!t]
\centering \caption{The first few values of $x_{n}, y_{n}, z_{n},
w_{n}$ on $X_n$.} \label{T1}
\begin{tabular}{ccccc}
\toprule[1pt]
$n$&0&1&2&3\\
\midrule
$x_{n}$&1&66&87,837,347&213,175,217,650,167,042,919,081,256\\

$y_{n}$&0&56&76,020,480&184,498,173,678,586,828,013,178,352\\

$z_{n}$&1&49&65,794,261&159,678,861,670,954,453,048,115,477\\

$w_{n}$&0&44&56,944,448&138,198,326,607,977,450,114,587,516\\

\bottomrule[1pt]
\end{tabular}
\end{table*}

\begin{table*}[!t]
\centering \caption{The first few values of $\alpha_n, \beta_n,
\gamma_n$ on $X_n$.} \label{T1}
\begin{tabular}{cccc}
\toprule[1pt]
$n$&1&2&3\\
\midrule
$\alpha_n$&0.8484848484848485&0.8654687623932904&0.8654766520813835\\

$\beta_n$&0.875&0.8654807362437070&0.8654766520839932\\

$\gamma_n$&0.8979591836734694&0.8654926301246852&0.8654766520866030\\
\bottomrule[1pt]
\end{tabular}
\end{table*}

By a similar argument as Lemma \ref{14}, the entropy for the number of dimer-monomers on $X_n$ is bounded by
\begin{align}
\frac{2\ln
x_{k}+\ln(\alpha_{n}^{2}+8\alpha_{n}+8)+2\ln(\alpha_{n}^{2}+2\alpha_{n}+2)}{7\times3^{k}}\leq
\mu_{X} \nonumber \\
\leq \frac{2\ln
x_{k}+\ln(\beta_{n}^{2}+8\beta_{n}+8)+2\ln(\beta_{n}^{2}+2\beta_{n}+2)}{7\times
3^{k}},\nonumber
\end{align}
where $k$ is a positive integer and
$\mu_{X}=\lim_{v(X_n)\rightarrow\infty}\frac{\ln m(X_{n})}{v(X_n)}$.
The convergence of the upper and lower bounds remains rapid.
More than a hundred significant figures for $X_n$ can be obtained when $k$
is equal to six. We have the following proposition.

\begin{prop}
The entropy per site for the number of dimer-monomers of
the graph $X_{n}$ in the large $n$ limit is
$\mu_{X}=0.6719549820008285\cdots$.
\end{prop}


\section{Appendix. Recursion relations for $X_{n}$ }

We give the recursion relations for the  graphs $X_{n}$ here. For
any positive integer $n$, we have\\

$x_{n+1} = T_{n}^{3}+3S_{n}^{2}T_{n}+6S_{n}T_{n}^{2}
+3x_{n}S_{n}^{2}+6x_{n}S_{n}T_{n}+6S_{n}^{3}+6x_{n}^{2}S_{n}+x_{n}^{3}
=32x_{n}^{3}+96x_{n}^{2}y_{n}+24x_{n}^{2}z_{n}+108x_{n}y_{n}^{2}+60x_{n}y_{n}z_{n}
+9x_{n}z_{n}^{2}+44y_{n}^{3}+39y_{n}^{2}z_{n}+12y_{n}z_{n}^{2}+z_{n}^{3}$,\\

$y_{n+1}
=T_{n}^{2}P_{n}+2S_{n}R_{n}T_{n}+S_{n}^{2}P_{n}+2R_{n}T_{n}^{2}+2S_{n}T_{n}P_{n}+2S_{n}T_{n}P_{n}
+2x_{n}S_{n}R_{n}+y_{n}S_{n}^{2}+2x_{n}R_{n}T_{n}+2S_{n}^{2}R_{n}+2S_{n}^{2}R_{n}+2y_{n}S_{n}T_{n}+2x_{n}S_{n}P_{n}
+2S_{n}^{2}R_{n}+2x_{n}y_{n}S_{n}+2x_{n}^{2}R_{n}+2x_{n}y_{n}S_{n}+x_{n}^{2}y_{n}
=32x_{n}^{2}y_{n}+32x_{n}^{2}z_{n}+8x_{n}^{2}w_{n}+64x_{n}y_{n}^{2}+88x_{n}y_{n}z_{n}
+20x_{n}y_{n}^{2}+20x_{n}z_{n}^{2}+6x_{n}z_{n}w_{n}+36y_{n}^{3}+64y_{n}^{2}z_{n}+13y_{n}^{2}w_{n}+29y_{n}z_{n}^{2}
+8y_{n}z_{n}w_{n}+4z_{n}^{3}+z_{n}^{2}w_{n}$,\\

$z_{n+1}=T_{n}P_{n}^{2}+2S_{n}R_{n}P_{n}
+R_{n}^{2}T_{n}+2S_{n}P_{n}^{2}+2R_{n}T_{n}P_{n}+2R_{n}T_{n}P_{n}
+2y_{n}S_{n}R_{n}+x_{n}R_{n}^{2}+2y_{n}S_{n}P_{n}+2S_{n}R_{n}^{2}+2S_{n}R_{n}^{2}+2x_{n}R_{n}P_{n}+2y_{n}R_{n}T_{n}
+2S_{n}R_{n}^{2}+2x_{n}y_{n}R_{n}+2y_{n}^{2}S_{n}+2x_{n}y_{n}R_{n}+x_{n}y_{n}^{2}
=32x_{n}y_{n}^{2}+64x_{n}y_{n}z_{n}+16x_{n}y_{n}w_{n}+36x_{n}z_{n}^{2}+20x_{n}z_{n}w_{n}
+z_{n}w_{n}^{2}+32y_{n}^{3}+80y_{n}^{2}z_{n}+20y_{n}^{2}w_{n}+64y_{n}z_{n}^{2}+32y_{n}z_{n}w_{n}
+4y_{n}w_{n}^{2}+13z_{n}^{3}+8z_{n}^{2}w_{n}+3x_{n}w_{n}^{2}$,\\

$w_{n+1}=P_{n}^{3}+3R_{n}^{2}P_{n}
+6R_{n}P_{n}^{2}+3y_{n}R_{n}^{2}+6y_{n}R_{n}P_{n}+6R_{n}^{3}+6y_{n}^{2}R_{n}+y_{n}^{3}
=32y_{n}^{3}+96y_{n}^{2}z_{n}+24y_{n}^{2}w_{n}+108y_{n}z_{n}^{2}+60y_{n}z_{n}w_{n}
+9y_{n}w_{n}^{2}+44z_{n}^{3}+39z_{n}^{2}w_{n}+12z_{n}w_{n}^{2}+w_{n}^{3}$.\\

  \begin{figure}[ht!]
\begin{center}
\includegraphics[width=14cm]{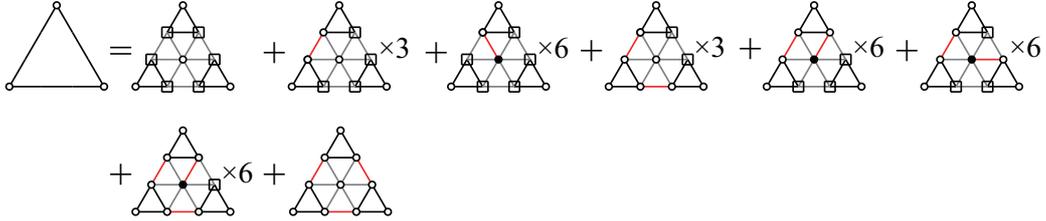}
\caption{Illustration for the expression of $x_{n+1}$.}
\end{center}
\end{figure}

  \begin{figure}[ht!]
\begin{center}
\includegraphics[width=14cm]{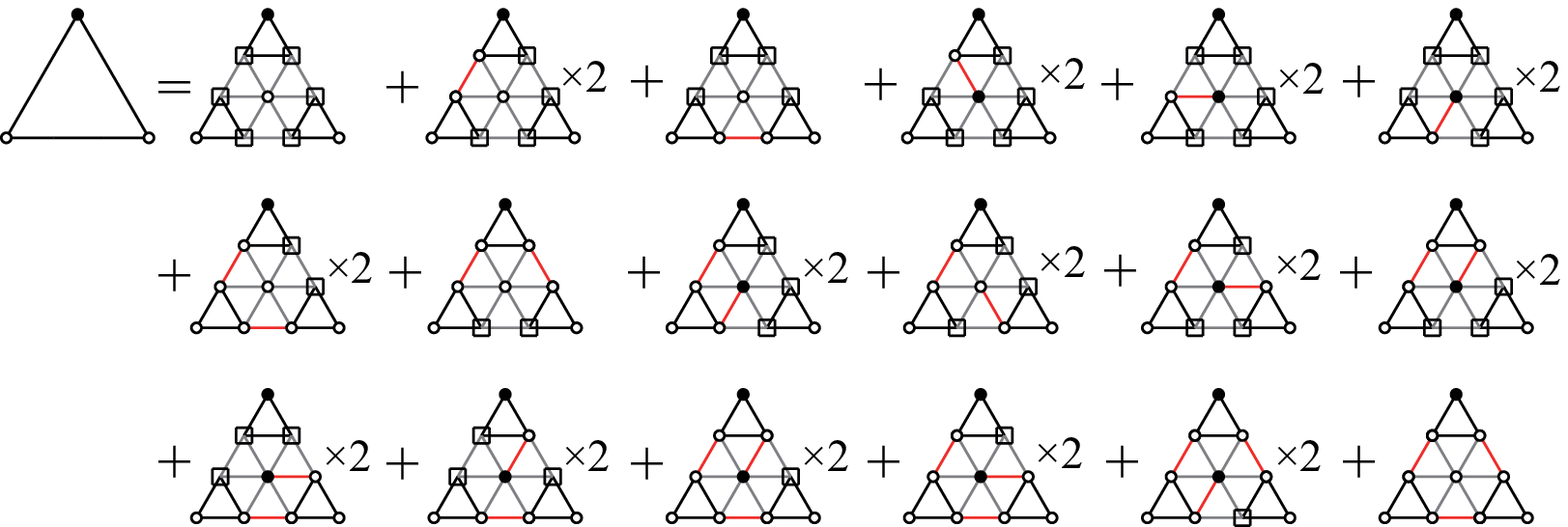}
\caption{Illustration for the expression of $y_{n+1}$. }
\end{center}
\end{figure}

  \begin{figure}[ht!]
\begin{center}
\includegraphics[width=14cm]{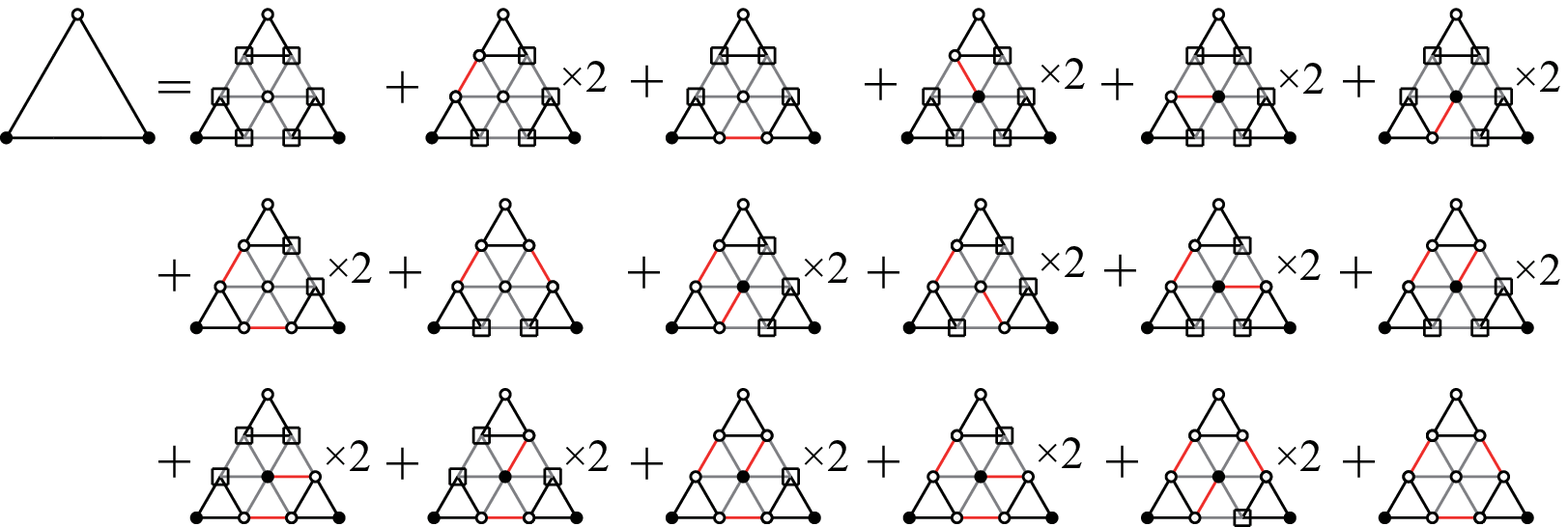}
\caption{Illustration for the expression of $z_{n+1}$.
}
\end{center}
\end{figure}

\begin{figure}[ht!]
\begin{center}
\includegraphics[width=14cm]{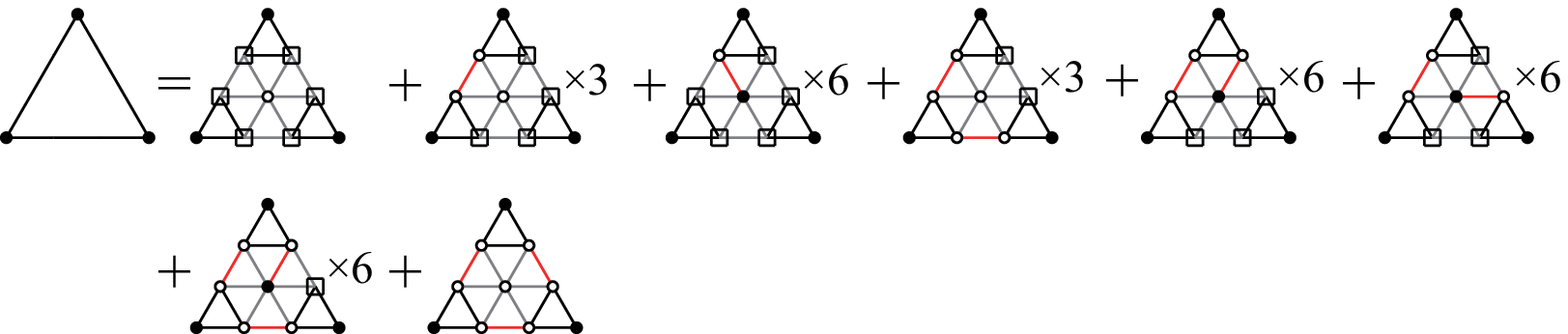}
\caption{Illustration for the expression of $w_{n+1}$.
}
\end{center}
\end{figure}

\end{document}